\documentclass[usenatbib]{mn2e}

\usepackage{graphicx}
\usepackage{natbib}
\def\citeN{\citet}
\def\cite{\citep}

\footnotesize
\newdimen\digitwidth    %define ! a one digit width for tables
\setbox0=\hbox{\rm0}
\digitwidth=\wd0
\catcode`!=\active
\def!{\kern\digitwidth}
\normalsize
\title{A high frequency search for radio pulsars in three EGRET error boxes}
\author[M.~J.~Keith et al.]
{M.~J.~Keith$^1$\thanks{Email: mkeith@pulsarastronomy.net}, S.~Johnston$^1$, M.~Kramer$^2$, P.~Weltevrede$^1$, K.~P.~Watters$^3$ and \newauthor
B.~W.~Stappers$^2$
\\
$^1$ Australia Telescope National Facility, CSIRO, P.O. Box 76, Epping, NSW 1710, Australia\\
$^2$ University of Manchester, Jodrell Bank Centre for Astrophysics, Alan Turing Building, Manchester M13 9PL, UK\\
$^3$ Department of Physics, Stanford University, Stanford, CA 94305, USA\\
}
\let\leqslant=\leq %Authors with AMS fonts and mssymb.tex can comment
\date{}
\begin{document}

\maketitle
\newcommand{\setthebls}{
%                 de-comment this line for double spacing:
%\baselineskip=20pt
}

\setthebls

\begin{abstract} 
We present a new survey for pulsars in the error boxes of the low-latitude
EGRET sources 3EG J1027$-$5817, 3EG J1800$-$2338 and 3EG J1810$-$1032.
Although all of these sources have been covered by previous pulsar surveys,
the recent discovery of the young, energetic pulsar PSR~J1410$-$6132 at
6.7~GHz has shown that pulsars of this type can be hidden from low
frequency surveys.
Using an observing frequency of 3.1~GHz we discovered a 91-ms pulsar,
PSR J1028$-$5819, which observations made at the Parkes telescope and
the Australia Telescope Compact Array have shown to be young and energetic.
We believe this pulsar is likely to be powering the unidentified
EGRET source 3EG J1027$-$5817. Like other energetic pulsars,
PSR J1028$-$5819 is highly linearly polarised, but astonishingly
has a pulse duty cycle of only 0.4\%, one of the smallest in the
entire pulsar catalogue.
\end{abstract}


\section{Introduction}
The EGRET telescope on board the Compton Gamma-Ray Observatory
detected several hundred gamma-ray sources, of which around half do not have a well established identification with any known object.
The true nature of these unidentified sources has been a matter 
for much debate (e.g. \citealp{hbb+99,kbm+03}) but the typically degree size error boxes of
the sources has make identification at other wavelengths difficult.
Pulsars are good candidates for many of the unidentified sources in
the Galactic plane, as they have a similar spatial distribution and 
are one of the few populations of astronomical objects 
positively identified as gamma-ray emitters. Indeed, a number of
recently discovered young pulsars are spatially coincident with unidentified
EGRET sources \cite{tbc01,kbm+03} but whether they pulse in gamma-rays
will only be determined following the launch of GLAST \cite{sgc+08}.

The recently discovered pulsar PSR J1410$-$6132 \cite{ojk+08}, which 
is coincident with a previously unidentified EGRET source, was recently detected as part of a
Galactic plane survey carried out at the high frequency of 6.5~GHz.
PSR~J1410$-$6132 was not discovered by surveys at lower frequencies because
it suffers from severe scatter broadening which renders the pulse invisible.
It is therefore sensible to assume that other Galactic-plane EGRET sources may be associated with pulsars that are not yet known due to obscuration from interstellar scattering.
To this end we carried out a survey of Galactic-plane EGRET sources using a high frequency receiver at the Parkes radio telescope.
Although PSR J1410$-$6132 was discovered using a 6.5~GHz multibeam
system, we decided to employ a wide-band 3~GHz receiver.
The frequency was chosen as a compromise between 
the interstellar scattering effects, which decrease as $\nu^{-4}$,
the typical pulsar spectral index (flux density decreases as $\nu^{-1.6}$, e.g. \citealp{lk05}),
and the number of pointings required to cover the large error boxes of
the EGRET sources.

In Section \ref{section_survobsan} of this paper we present our
survey of three unidentified galactic-plane EGRET sources.
In Section \ref{section_1028} we describe the discovery of the
young, energetic pulsar PSR J1028$-$5819. Finally in
Section \ref{section_disc} we discuss the emission properties
of J1028$-$5819 and the possible association between the
new pulsar and the EGRET source 3EG J1027$-$5817.

\section{Survey observations and analysis}
\label{section_survobsan}
\subsection{Observation strategy}
\begin{figure}
\includegraphics[width=6.5cm, angle=-90]{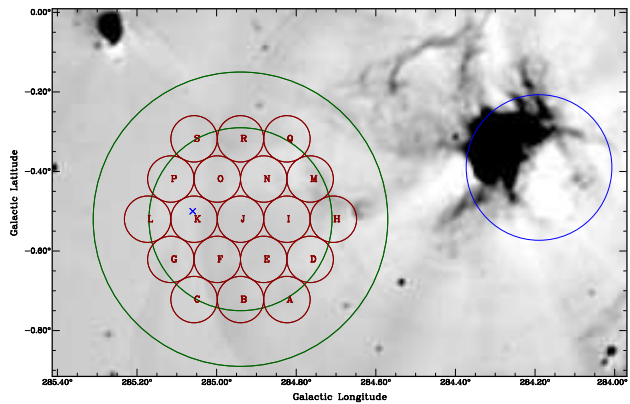}
\caption[]{
\label{1028map}
The pattern of 19 beams used to survey the error box of 3EG J1027$-$5817.
The 1-$\sigma$ and 2-$\sigma$ error circles are shown by the larger concentric circles.
The underlying image is an 843~MHz radio map from the Molonglo Galactic Plane Survey \cite{gcl98}.
The position of PSR J1028$-$5819 in beam K is marked with a cross and the position of the nearby HESS source J1023$-$575 \cite{aab+07} is marked with the blue circle to the right of the figure.
}
\end{figure}

We selected from the 3EG catalogue those sources with 
an absolute galactic latitude of less than 5\degr\ and a variability index, as discussed by \citeN{mmct96}, less than 2. This reduces the original
269 sources to just 31 and includes the known gamma-ray pulsars such
as Vela and PSR~B1706$-$44 and likely candidates such as
PSRs~B1046$-$58 and J1420$-$6048.
We then chose sources such that they were visible with the Parkes telescope
and that the error circle was less than $0.4^{\circ}$ in radius,
so that the 2-$\sigma$ error box could be filled with less than 20 pointings.
This resulted in our final selection of three sources:
3EG J1027$-$5817, 3EG J1800$-$2338 and 3EG J1810$-$1032.
As a confirmation, we repeated this process using the updated variability index suggested by \citeN{ntgm03} and produced the same three sources.
Each source was covered with 19 separate pointings; those for 3EG J1027$-$5817 are shown in Figure~\ref{1028map}.

Observations were carried out between 2008 April 6 and 9 at the Parkes radio telescope using the
10cm band of the dual `10--50' receiver at a centre observing frequency of 3.1~GHz.
This was connected to a 288 channel filterbank system with a total bandwidth of 864~MHz, 1-bit sampled in total intensity every 250 $\mu$s and
recorded to tape for off-line processing. Each beam
on the sky was observed for a total of 18 minutes.
The 10cm receiver has a system equivalent flux density of 45~Jy and
our sensitivity limit for pulsars with a 10\% duty cycle and low dispersion
measure was therefore 0.2~mJy.

%\begin{figure}
%\includegraphics[width=7.5cm]{egret2}
%\caption[]{
%\label{1800map}
%The pattern of 19 beams used to survey the error box of 3EG J1800$-$2338.
%The 1-$\sigma$ and 2-$\sigma$ error circles are shown by the larger concentric circles.
%The greyscale shows an L-band VLA image from the NVSS \cite{ccg+98}.
%}
%\end{figure}
%
%\begin{figure}
%\includegraphics[width=7.5cm]{egret3}
%\caption[]{
%\label{1810map}
%The pattern of 19 beams used to survey the error box of 3EG J1810$-$1032.
%The 1-$\sigma$ and 2-$\sigma$ error circles are shown by the larger concentric circles.
%The greyscale shows an L-band VLA image from the NVSS \cite{ccg+98}.
%}
%\end{figure}

\subsection{Analysis and results}
The recorded data were processed using software based on the {\sc Sigproc}\footnote{http://sigproc.sourceforge.net/} package.
Each pointing was dedispersed to 768 trial dispersion steps, up to a maximum of 1333 cm$^{-3}$pc.
Periodicities in the target file were then identified using a 
fast Fourier transform based algorithm.
The results were then collated into a number of candidates and optimised in the time domain using the {\sc PulsarHunter} software \cite{kei07}.
Candidates were then viewed using {\sc JReaper} (Keith~et~al., in preparation), to select the best candidates for re-observation.

This analysis resulted in one good candidate pulsar; immediate re-observation 
at Parkes confirmed that this was a genuine detection, and the new pulsar was 
designated PSR J1028$-$5819.

\section{PSR J1028$-$5819}
\label{section_1028}
PSR J1028$-$5819 was detected on 2008 April 6 in beam K (as labelled in Figure \ref{1028map}) with a dispersion measure of $96$~pc~cm$^{-3}$, during initial processing carried out at the telescope.
The detected pulse period of 91.4 ms immediately indicated that the 
pulsar was likely to be a young energetic pulsar and therefore potentially
associated with its EGRET source.
%Follow up observations confirmed the pulsar and also provided a first indication of the pulsar's spin down rate.

\subsection{Observations and results}
Observations of the pulsar were made on 2008 April 9
with the Australia
Telescope Compact Array (ATCA), an east-west synthesis telescope located
near Narrabri, NSW, which consists of six 22-m antennas on a 6 km track.
The observations were carried out simultaneously at
1.4 and 2.4~GHz with a bandwidth of 128~MHz at
each frequency subdivided into 32 spectral channels, and full Stokes
parameters. Total observation time on source was 120~min.
The ATCA is also capable of splitting each correlator cycle into
bins corresponding to different phases of a pulsar's period, and
in our case the pulse period of $\sim$91~ms was split into 32 phase bins.
This allows a search to be made for the pulsar over the field of view 
of the primary beam (40\arcmin) of the telescope.

Initial data reduction and analysis were carried
out with the \textsc{Miriad} package using
standard techniques. After flagging bad data, the primary calibrator
(PKS~1934$-$638) was used for flux density and bandpass calibration
and the secondary calibrator (PMN~J1051$-$5344) was used to solve for
antenna gains, phases and polarisation leakage terms.
After calibration, the data consist of 13 independent frequency channels
each 8~MHz wide for each of the 32 phase bins.
The data were then de-dispersed using the known dispersion measure.

From analysis of the interferometric data we determined the pulsar's position
to be at right ascension 10:28:$28.0\pm0.1$ and declination $-58$:19:$05.2\pm1.5$ in the J2000 coordinate system.
The pulsar has a flux density of $0.36\pm0.06$ and $0.52\pm0.06$~mJy
at 1.4 and 2.4~GHz respectively.
The pulsar therefore appears to have a rather flat spectral index, but we caution that this single flux measurement could be affected by scintillation.

Further observations of the pulsar were carried out using the 
Parkes radio telescope at $1.4$ and 3~GHz using the centre beam of
the 20cm-multibeam and 10--50 receivers respectively.
A digital filterbank was used to record integrated profiles with
full Stokes information every 30~s
with 1024 channels over a 256 or 1024 MHz band at $1.4$ and 3~GHz respectively.
In parallel, the data were sampled every 250~$\mu$s in total intensity using the 512 channel analogue filterbank.

The data were then further processed using standard pulsar timing techniques to yield 20 good time-of-arrival (TOA) measurements and integrated polarisation profiles.
The pulsar spin frequency, frequency derivative and dispersion measure were then fit to the TOAs using the {\sc TEMPO2} software package whilst holding the pulsar's position fixed.
From the timing analysis, we constrain the pulsar parameters as shown in Table \ref{params}.
This implies a characteristic age of 90~kyr, surface magnetic field strength of $1.2 \times 10^{12}$~G and a spin down energy rate of $8.3 \times 10^{35}$~erg~s$^{-1}$.
Using the measured dispersion measure and the \citeN{cl02} electron-density model, we derive a distance to the pulsar of 2.3~kpc.

\begin{table}
\caption[Timing parameters obtained for PSR J1027$-$5819.]{\label{params}
Timing parameters obtained for PSR J1027$-$5819.
Figures in parentheses are the nominal 1$\sigma$ \textsc{tempo2} uncertainties in the least-significant digits quoted.
}
\begin{tabular}{ll}
\hline
Pulsar name\dotfill & J1028$-$5819 \\ 
 Right ascension, $\alpha$\dotfill & 10:28:28.0 \\
  Declination, $\delta$\dotfill & $-$58:19:05.2 \\
 Pulse period, $P$ (ms)\dotfill & 91.4032309(14)\\
Period derivative, $\dot{P}$ \dotfill & 16.1(8)$\times 10^{-15}$\\
Pulse frequency, $\nu$ (s$^{-1}$)\dotfill & 10.94053230(16) \\
Frequency derivative, $\dot{\nu}$ (s$^{-2}$)\dotfill & $-$1.92(9)$\times 10^{-12}$ \\
  Dispersion measure, DM (cm$^{-3}$pc)\dotfill & 96.525(2) \\
   Epoch of fit (MJD)\dotfill & 54562 \\
\\

 $\log_{10}$(Characteristic age, yr) \dotfill & 4.95 \\
  $\log_{10}$(Surface $B$ field strength, G) \dotfill & 12.09 \\
    $\log_{10}$(Energy loss rate, $\dot{E}$) \dotfill & 35.91 \\
      DM derived distance (kpc)\dotfill & 2.3\\
     \\

MJD range of fit\dotfill &    54563.6---54606.2 \\ 
Number of TOAs\dotfill & 20 \\
Rms timing residual ($\mu s$)\dotfill & 4.5 \\
 Clock correction procedure\dotfill & TT(TAI) \\
  Solar system ephemeris model\dotfill & DE405 \\
 \hline
 \end{tabular}

\end{table}

\subsection{Polarisation profiles and single pulses}
\begin{figure}
\includegraphics[height=8cm,angle=-90]{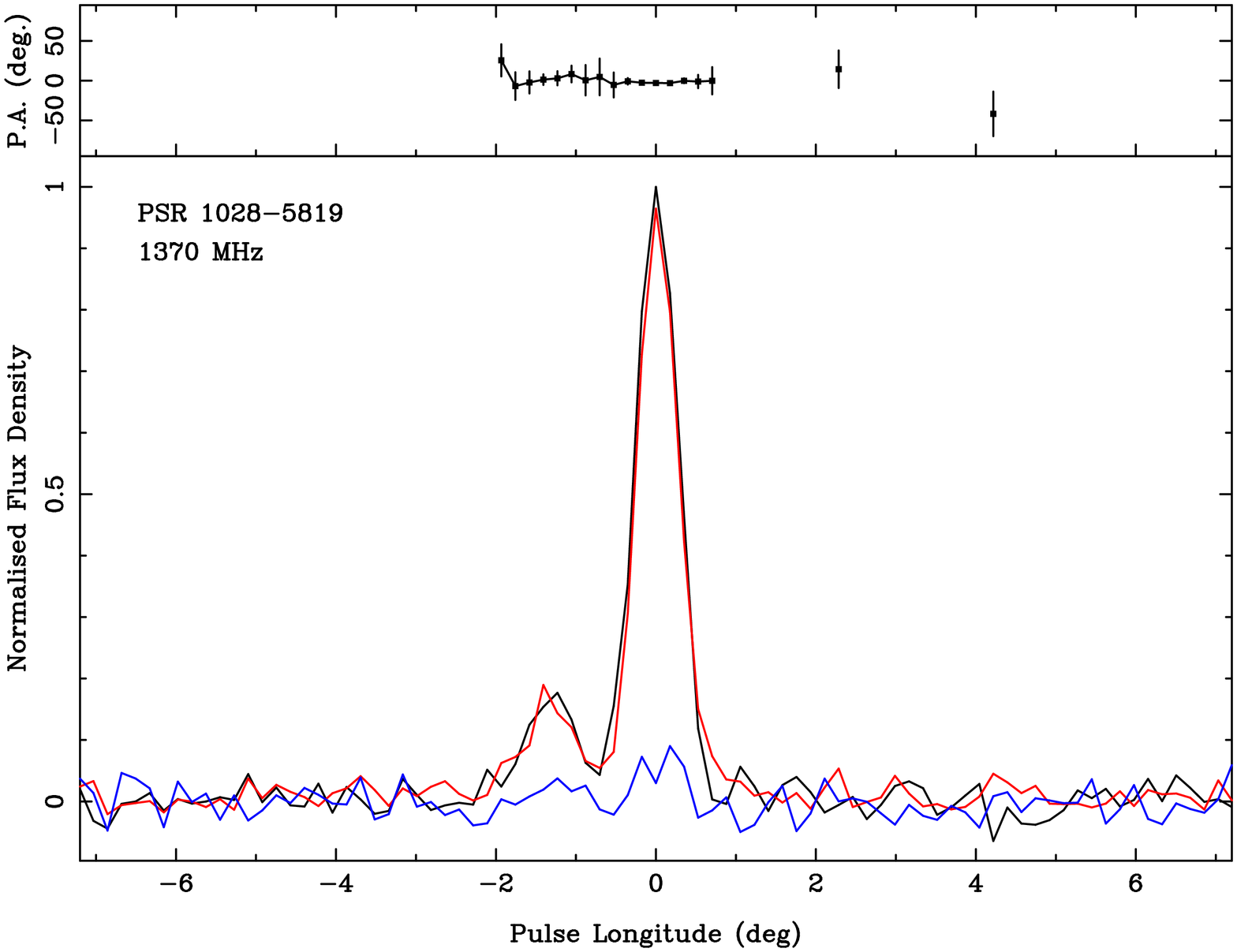}
\includegraphics[height=8cm,angle=-90]{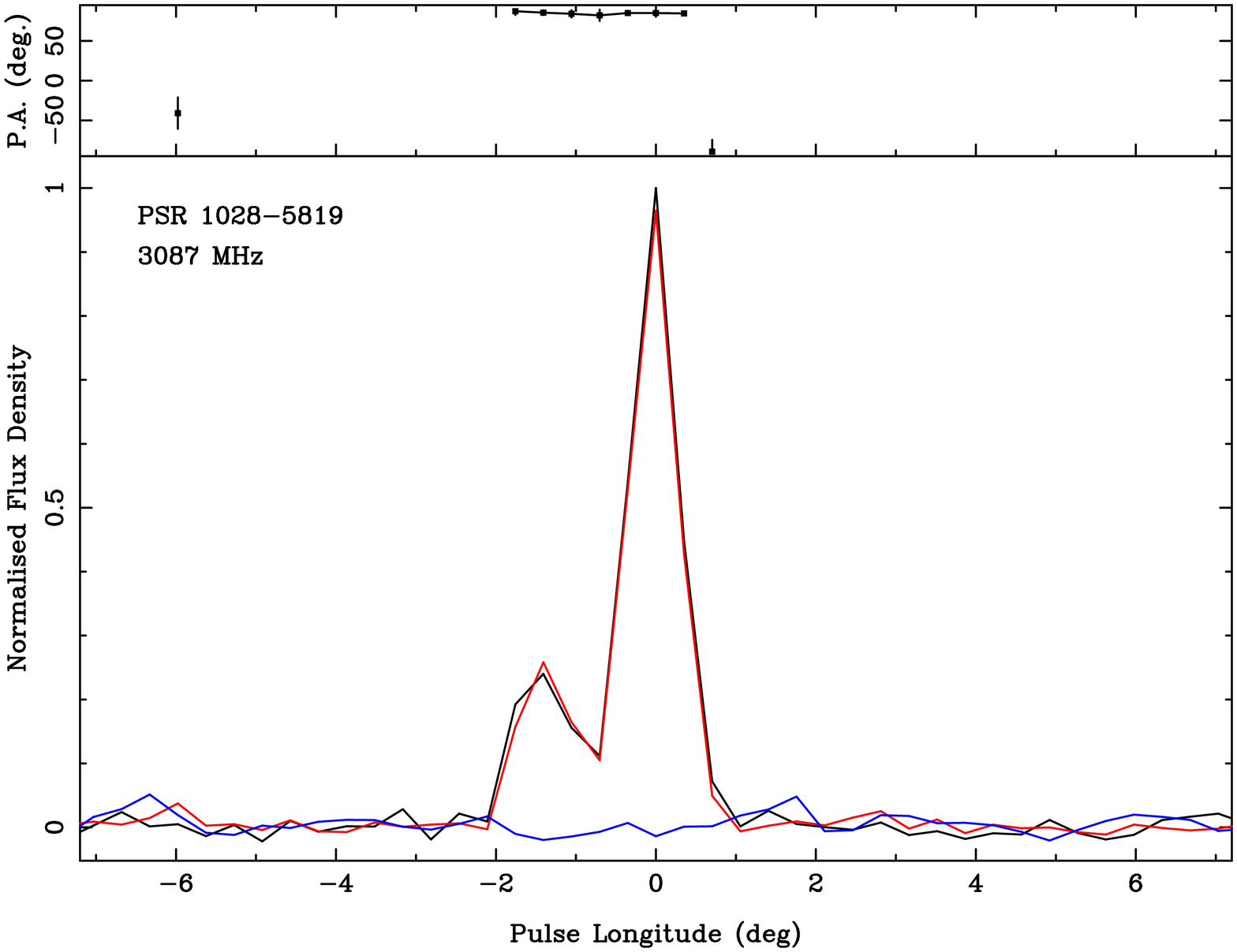}
\caption[]{
\label{1400prof}
Integrated pulse profile of J1028$-$5819 at 1.4 GHz (top) and 3.1~GHz (bottom).
Total intensity is shown in black, with linear polarisation in red and circular polarisation in blue.
The polarisation position angle variation over the pulse is shown in the upper panel.
The polarisation position angles shown are not absolute.
The 1.4~GHz profile is generated by summing four 4.5-minute observations taken at Parkes on 2008 April 15 with 2048 bins over the entire profile.
The 3.1~GHz profile is produced from a single 6-minute observation taken at Parkes on 2008 April 9, with 1024 bins across the profile.
}
\end{figure}

The integrated profiles for PSR~J1028$-$5819 at 1.4~GHz and 3.1~GHz 
are shown in Figure \ref{1400prof}.
We were unable to detect the pulsar during a 10~min observation at 0.61~GHz, likely
due to a combination of the pulsar's relatively flat spectral index, the increased
sky brightness temperature at this low frequency and radio interference.
We used the 20~cm observations to determine the rotation measure of the pulsar
to be $-5 \pm 4$ rad m$^{-2}$.
%TODO: talk to Aris: CHECK FOR CONSISTENCY AND DISTANCE

The profiles show that the pulse is a narrow double, with the leading component
weaker than the trailing component but with a flatter spectral index. Both
components are virtually 100 per cent linearly polarised and there is little
circular polarisation.
These characteristics are typical of young, energetic pulsars
(e.g. \citealp{jw06} and Weltevrede \& Johnston, in preparation).
The position angle is almost completely flat across the
entire profile.

Unlike other energetic pulsars, the width of the profile of PSR~J1028$-$5819
is remarkably narrow.
The two components have a half-power width of only 175~$\mu$s and 160~$\mu$s for the leading and trailing peaks respectively.
The separation of the components is 330~$\mu$s and the 10$\%$ power width of the entire profile is 560~$\mu$s.
This duty cycle of just 0.4$\%$ appears to be the smallest of any integrated profiles of the known pulsars.

We note that this small duty cycle is akin to that of the RRATs \cite{mll+06}
which show very bright pulses followed by long period nulls.
\citeN{jr02} and \citeN{wsrw06} showed that some so-called normal
pulsars often showed narrow, bright pulses quite distinct from the normal
single pulses. We therefore examined 6500 single pulses from PSR~J1028$-$5819
to look for evidence of similar behaviour. However, the pulsar's flux does not appear to be strongly modulated and we detected no
single pulses with a flux density greater than 10 times the mean on-pulse 
flux density.

%\begin{figure}
%\includegraphics[height=8cm,angle=-90]{20cm.ps}
%\caption[]{
%\label{1400prof}
%Integrated pulse profile of J1028$-$5819 at 1.4 GHz.
%Total intensity is shown in black, with linear polarisation in red and circular polarisation in blue.
%The polarisation position angle variation over the pulse is shown in the upper panel.
%This profile is generated by summing four 4.5-minute observations taken at Parkes on 2008 April 15 with 2048 bins over the entire profile.
%}
%\end{figure}
%
%\begin{figure}
%\includegraphics[height=8cm,angle=-90]{10cm.ps}
%\caption[]{
%\label{3000prof}
%Integrated pulse profile of J1028$-$5819 at 3.094 GHz.
%Total intensity is shown in black, with linear polarisation in red and circular polarisation in blue.
%The polarisation position angle variation over the pulse is shown in the upper panel.
%This profile is produced from a single 6-minute observation taken at Parkes on 2008 April 9, with 1024 bins across the profile.
%}
%\end{figure}

\section{Discussion}
\label{section_disc}

\subsection{Association with 3EG J1027$-$5817}
The pulsar dispersion measure can be used to estimate the distance to the source by using a model of the galaxy electron density.
Using the model of Cordes \& Lazio (2002)\nocite{cl02}, the estimated distance 
to the pulsar is 2.3 kpc.

The spin down rate of PSR J1028$-$5819 implies a rotational energy loss rate $\dot{E} = 8.3 \times 10^{35}$~erg~s$^{-1}$.
At a distance of 2.3 kpc we compute the value $\dot{E}/d^2$, a good indicator of gamma-ray detectability, to be $1.6 \times 10^{35}$~erg~s$^{-1}$~kpc$^{-2}$, which is in the top 20 of all known pulsars.

By fixing the distance to the EGRET source at the pulsar distance we can compute the power output of the gamma-ray source.
This distance plus the published photon count rate of $6.6 \times 10^{-7}$~photons~cm$^{-2}$~s$^{-1}$ and photon spectral index of 2 for 3EG J1027$-$5817 \cite{hbb+99} implies a gamma-ray energy-loss rate of $2.5 \times 10^{34}$~erg~s$^{-1}$, with an assumed beaming angle of $1/(4\pi)$.
This implies that if the pulsar is powering the EGRET source then $\sim 3\%$ of the pulsar spin down energy must be going into gamma-ray flux.
This is consistent with the percentage seen in other gamma-ray pulsars (as discussed by e.g. \citealp{ojk+08}) and we therefore claim that PSR~J1028$-$5819 is indeed the source responsible for the gamma-rays from 3EG~J1027$-$5817.

Confirmation or rejection of this hypothesis will be possible with the upcoming GLAST mission.
This will be able to both confirm the nature of the EGRET source and to further measure the gamma-ray properties of the pulsar.
In order to provide accurate rotational ephemeris for gamma-ray folding, we will continue to monitor the pulsar in the radio band.

\subsection{Multi-wavelength data and possible associations}
As shown in Figure \ref{1400prof}, PSR J1028$-$5819 is easily detectable at 1.4 GHz with minimal scattering effects.
Since its position is within the bounds of the Parkes Multibeam Pulsar Survey \cite{mlc+01}, an investigation was carried out to determine why this pulsar was not discovered in the original survey.
Inspection of the observation logs and data archives shows that technical issues meant that data from the survey observation covering the pulsar were never processed.
We obtained these data, taken on 2004 October 10, and the pulsar was indeed detected with a signal-to-noise ratio of 25.

In the X-ray band, data from the ROSAT All-Sky Survey does not show any significant emission at the pulsar position.
However observations with more sensitive instruments should be undertaken to determine whether the pulsar is an X-ray emitter and/or uncover the presence of an underlying X-ray pulsar wind nebula.

PSR J1028$-$5819 is located just $1\degr$ away from the HESS source J1023$-$575, although the latter is likely associated with the stellar cluster Westerlund 2 \cite{aab+07}.
However, we can consider the possibility that the HESS source was powered by PSR~J1028$-$5819 sometime in the past.
If the pulsar was born at the HESS source position, using the DM derived distance and the characteristic age as true values, we compute a transverse velocity of $\sim500$~km~s$^{-1}$.
This is somewhat high, but not outside of the observed range of known pulsar velocities.
The distance to Westerlund 2 is not well known, although most estimates put it at around 8 kpc (e.g. \citealp{rmg+07}) which is considerably further than our estimated distance to the pulsar.
At this time we do not have any evidence to suggest that PSR J1028$-$5819 is related to HESS J1023$-$575 or Westerlund 2, however the small angular separation should be noted.

\subsection{Implications of small duty cycle}
Like many other young, energetic pulsars, PSR~J1028$-$5819 is highly
polarised and shows a double peaked profile (e.g. Weltevrede \& Johnston, in preparation).
However, the pulsar has an unusually small duty cycle, more than an
order of magnitude smaller than the norm. How does this fit into
our picture of pulsar beams?

Since the profile is double peaked with similar pulse widths for
both leading and trailing parts, we can hypothesise that we are cutting through a cone of emission, as in the typical pulsar emission scenario.
If we assume that the visible emission represents the entire opening 
angle of the open field region of the pulsar magnetosphere, we
compute an emission height of only 2~km for the 1.4~GHz emission.
This is below the theoretical radius for a neutron star and strongly suggests that this simple model cannot be applied in this case.

An alternative explanation comes from assuming a more conventional 
emission height of a few hundred km, giving a beam opening angle
of $\sim 20\degr$. In this case, only a small fraction of the beam
happens to be illuminated, resulting in a narrow observed profile
or we happen to be just grazing the very outer edge of the beam.
In this case it is difficult to explain the double-peaked nature of the profile.
If the observed pulsations are due to a grazing geometry and we assume radius-to-frequency mapping \cite{cor78}, we might expect that the pulsar
would disappear at high frequencies as our line of sight no longer
intersects the beam.
The evolution of profile with frequency will provide valuable insight, therefore observations at other radio frequencies are warranted.

This discovery highlights the limits in our understanding of pulsar beams and shows that we may need to update our beaming models and detection probabilities for young pulsars.

\section{Conclusions}
Observations of three EGERT sources 3EG J1027$-$5817, 
3EG J1800$-$2338 and 3EG J1810$-$1032 at 3.1~GHz resulted in the
detection of a previously unknown young and energetic pulsar,
PSR~J1028$-$5819. The parameters of the pulsar lead us to conclude that
it is likely to be the source powering 3EG J1027$-$5817 and we confidently
expect that GLAST will detect gamma-ray pulsations from the pulsar
in the near future. The pulsar is highly polarised and has an 
unusually narrow pulse profile.

\section*{Acknowledgements}
This research was partly funded by grants from the Science \& Technology Facilities Council, UK. The Australia Telescope
is funded by the Commonwealth of Australia for operation as a National
Facility managed by the CSIRO.
We thank P. Edwards and R. Wark for very prompt allocation 
of target of opportunity time at the Australia Telescope Compact Array
and J. Reynolds and J. Sarkissian for observations with Parkes.

\bibliographystyle{mnras}
\bibliography{journals,myrefs,psrrefs,modrefs}

\end{document}